\documentclass[%
 pre,
rsi,%
 amsmath,amssymb,
 reprint,%
superscriptaddress
]{revtex4-1}

\usepackage{graphicx}
\usepackage{dcolumn}
\usepackage{bm}
\usepackage{color}

\begin{document}

\def\bea{\begin{eqnarray}}
\def\eea{\end{eqnarray}}
\def\beq{\begin{equation}}
\def\eeq{\end{equation}}
\def\f{\frac}
\def\k{\kappa}
\def\e{\epsilon}
\def\ve{\varepsilon}
\def\be{\beta}
\def\D{\Delta}
\def\h{\theta}
\def\t{\tau}
\def\a{\alpha}

\def\cDa{{\cal D}[X]}
\def\cD{{\cal D}[x]}
\def\cL{{\cal L}}
\def\cLo{{\cal L}_0}
\def\cLa{{\cal L}_1}
\def\rv{{\bf r}}
\def\tv{{\bf {\mathfrak t}}}
\def\on{{\omega_{\rm on}}}
\def\off{{\omega_{\rm off}}}
\def\fv{{\bf{f}}}
\def\fm{\bf{f}_m}
\def\zh{\hat{z}}
\def\yh{\hat{y}}
\def\xh{\hat{x}}
\def\km{k_{m}}

\def\Re{{\rm Re}}
\def\sj{\sum_{j=1}^2}
\def\rk{\rho^{ (k) }}
\def\rek{\rho^{ (1) }}
\def\cek{C^{ (1) }}
\def\rz{\rho^{ (0) }}
\def\rt{\rho^{ (2) }}
\def\rtb{\bar \rho^{ (2) }}
\def\trk{\tilde\rho^{ (k) }}
\def\trek{\tilde\rho^{ (1) }}
\def\trz{\tilde\rho^{ (0) }}
\def\trt{\tilde\rho^{ (2) }}
\def\r{\rho}
\def\tD{\tilde {D}}

\def\s{\sigma}
\def\kb{k_B}
\def\bF{\bar{\cal F}}
\def\F{{\cal F}}
\def\la{\langle}
\def\ra{\rangle}
\def\nn{\nonumber}
\def\up{\uparrow}
\def\dn{\downarrow}
\def\S{\Sigma}
\def\dg{\dagger}
\def\d{\delta}
\def\p{\partial}
\def\l{\lambda}
\def\L{\Lambda}
\def\G{\Gamma}
\def\o{\Omega}
\def\w{\omega}
\def\g{\gamma}

\def\vv{ {\bf v}}
\def\jv{ {\bf j}}
\def\jr{ {\bf j}_r}
\def\jd{ {\bf j}_d}
\def\jdd{ { j}_d}
\def\noi{\noindent}
\def\a{\alpha}
\def\d{\delta}
\def\p{\partial} 

\def\la{\langle}
\def\ra{\rangle}
\def\e{\epsilon}
\def\n{\eta}
\def\g{\gamma}
\def\break#1{\pagebreak \vspace*{#1}}
\def\hf{\frac{1}{2}}
\def\rcurs{r_{ij}}

\def\bv{ {\bf b}}
\def\uv{ {\hat u}}
\def\rv{ {\bf r}}

\title{Morphological and dynamical properties of semiflexible filaments driven by molecular motors}

\author{Nisha Gupta}
\email{nishagupta@iisermohali.ac.in}
\author{Abhishek Chaudhuri}
\email{abhishek@iisermohali.ac.in}
\affiliation{Indian Institute of Science Education and Research Mohali, Knowledge City, Sector 81, SAS Nagar - 140306, Punjab, India.}
\author{Debasish Chaudhuri}
\email{debc@iopb.res.in}
\affiliation{Institute of Physics, Sachivalaya Marg, Bhubaneswar 751005, India.} 
\affiliation{Homi Bhaba National Institute, Anushaktigar, Mumbai 400094, India.}

\date{\today}

\begin{abstract}
We consider an explicit model of a semiflexible filament moving in two dimensions on a gliding assay of motor proteins,  which attach to and detach from  filament segments stochastically, with a detachment rate that depends on the local load experienced. Attached motor proteins move along the filament to one of its ends with a velocity that varies non-linearly with the motor protein extension. The resultant force on the filament drives it out of equilibrium. The distance from equilibrium is reflected in the end-to-end distribution, modified bending stiffness, and a transition to spiral morphology of the polymer. The local stress dependence of activity results in correlated fluctuations in the speed and direction of the center of mass leading to a series of ballistic-diffusive crossovers in its dynamics. 
\end{abstract}

\maketitle

\section{Introduction}
The active cytoskeleton in a living cell provides its structural stability, mediates deformation and growth of the cell when necessary, and acts as transport lanes and highways for intracellular cargo~\cite{Howard2005, Alberts2009}. It is made of semiflexible filaments, e.g., F-actins and microtubules, that are driven by associated motor proteins, for example myosin and kinesin family of motor proteins respectively~\cite{Vale2003, Chowdhury2013, Hancock1999, Fletcher2010, Huber2013a, MacKintosh1995, Yamaoka2012, Kreten2018}. Given the complexity of the cytoskeleton in  a living cell, {\em in vitro} experiments were devised in which purified and stabilized cytoskeletal filaments and corresponding motor proteins were studied separately~\cite{Kron1986, Sekimoto1995, Surrey2001, Schaller2010, Sumino2012, Bourdieu1995c,Kierfeld2008b}. 
Single molecule experiments on motor proteins revealed details of their dynamics, e.g., force-velocity relation, dependence of turnover on load experienced, and dependence of activity on ATP concentration~\cite{Oiwa1990a, Hancock1999, Coy1999a, Mehta1999, Rief2000a, Schnitzer2000,  Rock2001a, Yildiz2003}. Motion of rigid cargo under collective drive of molecular motors has been studied both experimentally and theoretically~\cite{Gross2002, Kural2005, Julicher1995, Vilfan1998, Badoual2002, Grill2005,Klumpp2005,Leduc2010b,Kraikivski2006, Nair2016}. In a gliding assay setup, heads of molecular motors are attached to a suitably prepared cover slit irreversibly, such that the tails can actively drive the associated filaments, hydrolyzing the chemical fuel ATP. This led to observation of collective motion, e.g., formation of spiral and aster patterns in microtubules driven by kinesin~\cite{Surrey2001,Ndlec1997} or dynein molecules~\cite{Sumino2012}, or swirling patterns in high density F-actins floating on a myosin motility assay~\cite{Schaller2010}.  

\begin{figure}[t]
\begin{center}
\includegraphics[width=8.5cm]{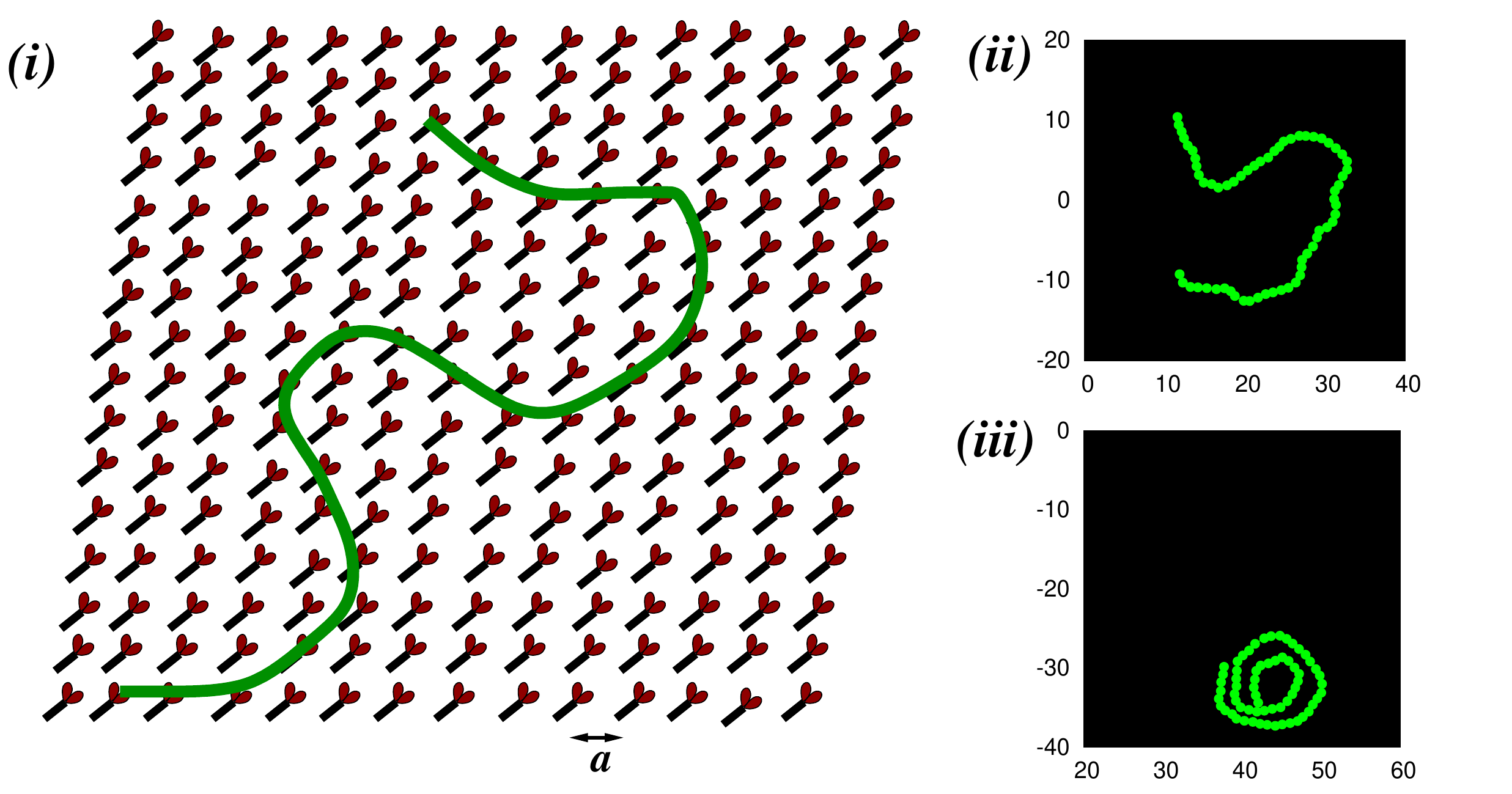}
	\caption{ (color online) (a) Schematic of the system showing the molecular motors arranged on a square grid. The semiflexible polymer glides on the bed of molecular motors. (b) and (c) Simulation snapshots of the polymer in an open and spiral state for a polymer with persistence ratio $u=3.33$, under the influence of MP activity $Pe=100$, and bare persistence ratio $\o=5/6$.
}
\label{figschem}
\end{center}
\end{figure}

Such patterns were explained within active hydrodynamics framework, and agent based models~\cite{Lee2001, Sankararaman2004, Sumino2012, Schaller2010}. Spiral rotation and flagella like beating of individual filaments were reproduced within effective active polymer models, modeling activity as a tangential self-propulsion~\cite{Sekimoto1995,Bourdieu1995c,Jiang2014a, Jiang2014, Isele-Holder2015, Chelakkot2014a}, stresslets distributed over the filament contour~\cite{Jayaraman2012,Laskar2013}, or chemical activity~\cite{Sarkar2017, Sarkar2016, Sarkar2014}, in presence or absence of hydrodynamic coupling. 
The collective dynamics in such models change from  coherently free flowing motion to frozen spiraling ones with changing activity~\cite{Prathyusha2018, Duman2018}.   
Generic consideration of a stiff filament in an active medium leads to the possibility of both increase or decrease of effective bending rigidity, depending on the orientation of filament segments with respect to contractile or extensile medium~\cite{Kikuchi2009, Gladrow2016a}.  
It was shown that a semiflexible filament under active correlated noise transform from bending rigidity dominated to flexible polymer- like dynamics~\cite{Eisenstecken2016}.  Their center of mass motion showed a single crossover from a short time ballistic, to long time diffusive behavior~\cite{Isele-Holder2015, Ghosh2014}. In contrast, as we show in this paper, a more microscopic consideration of both the cytoskeletal filament and motor proteins allows for local stress relaxation leading to novel behavior, e.g.,  a series of ballistic- diffusive crossovers of the filament center of mass.

{ 
Previous studies either modeled the motor proteins explicitly considering the driven object as a rigid cargo, or modeled the mechanical properties of the driven polymer explicitly, using self propulsion devoid of any underlying mechanism for relaxation.
Thus the impact of stress dependent dynamics of motor proteins on the filament properties, despite its importance, remains elusive within such models. In this paper we set out to address this issue. 
We perform numerical simulations, explicitly modeling the mechanical properties of the filament, and that of individual motor proteins as active harmonic springs undergoing attachment- detachment kinetics that do not obey detailed balance. The  attachment to filament is diffusion limited, and the detachment rate increases exponentially with the extension of individual motor proteins. In the attached state the {\em tail} of a motor protein moves tangentially towards one end of the polymer in an active manner, with a velocity that depends non-linearly on the motor protein extension.

We characterize the non-equilibrium conformations of the polymer comparing its end-to-end distribution with that of the equilibrium filament. In theoretical studies of active systems, key concepts such as broken detailed balance and entropy production has recently been used to characterize the distance of these systems from their equilibrium counterparts~\cite{Gladrow2016a, Fodor2016, Ganguly2013, Chaudhuri2014}. We show that subtle changes in the local load dependence of detachment rate and active velocity of motor proteins, leads to dramatic difference in the end-to-end distribution. With increasing activity, the difference increases,  the effective bending stiffness reduces, and the polymer shows a phase coexistence between open and spiral chains.
The most startling result is seen in the dynamics. The center of mass of the polymer shows a series of crossovers between ballistic and diffusive motion, controlled by its inertial, orientational and speed relaxation times scales.

In Sec.\ref{sec:model} we present the model and  details of the numerical simulation. All the results are discussed and analysed in Section~\ref{results}.  Finally, we present a summary and outlook in Sec.~\ref{sec:outlook}.

\begin{figure*}[t]
\begin{center}
\includegraphics[width=18cm]{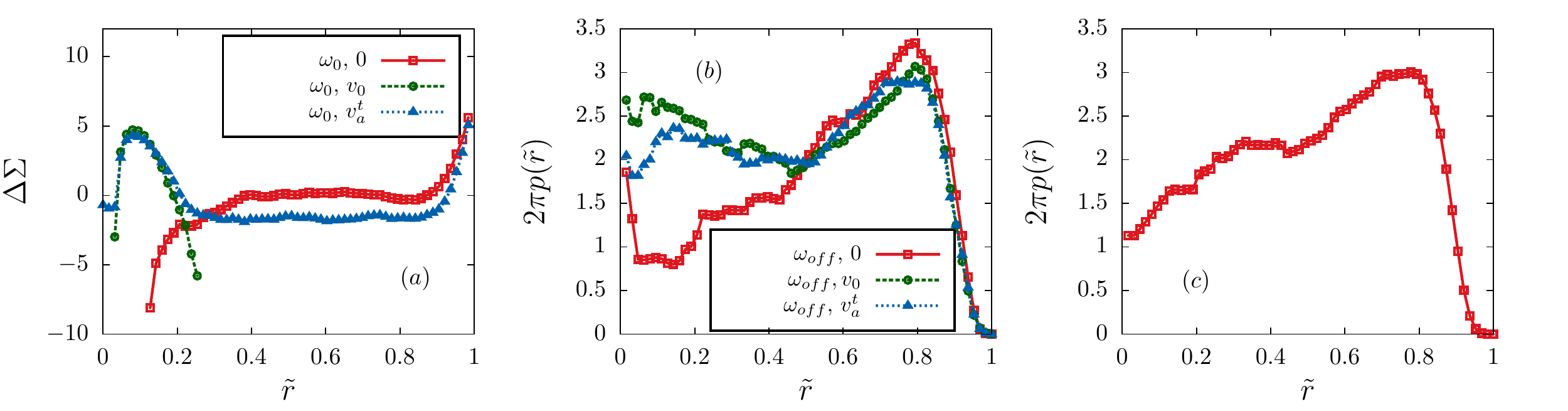}
\caption{ (color online) Activity dependence of end-to-end distribution functions and their difference from equilibrium for a filament with $N=64$ having persistence ratio $u=3.33$. The MP activity is controlled by turnover with a bare processivity $\o = 5/6$, and  non-zero active velocity $v_0$ set by $Pe=1$. ($a$)\,The logarithmic ratio of probabilities of filament under active drive with respect to that of the equilibrium polymer, $\Delta\Sigma$, provides a measure of the difference in distributions.  The legends denote parameter values (detachment rate, MP velocity), where, in this figure, all data sets correspond to a constant detachment rate $\w_0$, and MP velocity varies between constant values $0$, $v_0$, and stretching dependent active velocity $v_t^a$ as denoted by Eq.(\ref{velc}). ($b$)\,The end-to-end distribution of stretchable semiflexible polymer $p(\tilde r)$, { with local strain dependent detachment $\off$ as in Eq.\ref{eq:off}. ($c$)~The end-to-end distribution of the stretchable semiflexible polymer at equilibrium $p_{eq}(\tilde r)$.}   
}
\label{figKL}
\end{center}
\end{figure*}

\section{Model and Simulations}
\label{sec:model}
We consider an extensible semiflexible filament described as a bead-spring chain of $N$ beads constituting $(N-1)$ bonds of equilibrium length $\s$ such that the chain length $L=(N-1)\s$, spring constant $A$, and finite bending rigidity $\k$. This is described by  the Hamiltonian
\bea
\be H = \sum_{i=1}^{N-1} \f{A}{2 \s} \left[ \bv (i) - \s \tv(i)  \right]^2 + \sum_{i=1}^{N-2} \f{\k}{2 \s} \left[ \tv (i+1) - \tv(i)  \right]^2,\nn\\
\eea
with $\be = 1/\kb T$, the inverse temperature.
The bond vector $\bv(i)=\rv(i+1)-\rv(i)$, where $\rv(i)$ denotes the position of the $i$-th bead. This allows one to define the local tangent $\tv(i) = [\rv(i+1)-\rv(i)]/b(i)$.
In the limit of large $A$,  instantaneous bond lengths $b(i) \approx \s$, and the polymer maps to a worm like chain~\cite{Dhar2002}. In addition, excluded volume interactions between the non-bonded beads of the polymer is incorporated via a Weeks-Chandler-Anderson (WCA) potential
$\be V_{WCA}(\rcurs) = 4[(\s/\rcurs)^{12} - (\s/\rcurs)^6 + 1/4]$ if $\rcurs < 2^{1/6}\s$ and $0$ otherwise.

The polymer is placed on a substrate of motor protein (MP) assay (Fig.~\ref{figschem}($a$)\,). We explicitly model MPs and their dynamics, unlike several recent studies that used effective active polymer models~\cite{Eisenstecken2016,  Isele-Holder2015, Yang2010, Ghosh2014}. The MPs are modeled as {\em active} elastic linkers. We assume the MP heads are attached irreversibly to the substrate at position $\rv_0^i=(x_0^i,y_0^i)$ placed on a two dimensional square lattice with lattice parameter $a$ determined by the MP density $\r$. The polymer floats on this MP bed.  The {\em tails} of MPs may bind (unbind) to (from) the nearest polymer segments stochastically. 
The attachment process is diffusion limited. The tail of a MP attaches to a polymer segment if it lies within a capture radius $r_c$ with an attachment rate $\on$. The extension $\Delta \rv$ of the MP in the attached state generates an elastic load $\fv_l = -k_m\Delta{\bf r}$. An attached MP unbinds from a polymer segment with a rate $\off$ which depends on the stress felt by the MP as 
\bea 
\off = \omega_0\exp(f_l/f_d),
\label{eq:off}
\eea 
where $\omega_0$ is the bare off rate, $f_l = |\fv_l|$ and $f_d$ is the detachment force. The ratio $\on:\off$ does not obey detailed balance. When attached, a MP can move on the filament towards one of its ends, depending on the MP and filament type. For example, attached Kinesin moves towards positive end of the microtubule with {\em active} velocity $v^a_t$ along the local tangent of the filament given by~\cite{Schnitzer2000, Chaudhuri2014c}
\bea
v^a_t(f_t) = \f{v_0}{1+d_0\exp(f_t/f_s)},
\label{velc} 
\eea
where $f_t=-\fv_l.\tv$, $d_0 = 0.01$ and $f_s$ is the stall force. Here $v_0$ denotes the velocity of MP in the absence of stress. The extension of a given MP depends on the duration and velocity with which it moves along the filament before detachment, as well as the movement of the filament segment it is attached to. This generates a stochastic and non-uniform elastic load on different MPs.

We perform numerical simulations of the model to investigate structural and dynamical properties of the polymer, actively driven by MPs. The molecular dynamics of polymer is performed using the velocity-Verlet algorithm in presence of a Langevin heat bath. The bath fixes the ambient temperature $\kb T$ through a Gaussian white noise obeying $\la \eta_i(t) \ra =0$, and $\la \eta_i(t) \eta_j (t') \ra = 2 \a \kb T \d_{ij} \d(t-t')$, where $\a = 3 \pi \eta \s$ with $\eta$ denoting viscosity of the environment. This defines the diffusivity over the bead size $\s$, $D = k_BT/\alpha$. 
The units of energy, length and time are set by $\kb T$,   $\sigma$ and $\tau = \alpha \sigma^2/k_B T$, respectively. 

We set out to perform numerical simulations to study conformational and dynamical properties involving the longest length and time scales of the polymer, under the influence of an active MP bed pumping energy from the shortest length scales. The large separation between length and time-scales makes a fully microscopic parametrization of molecular motors prohibitively expensive in terms of simulation time. For example, the capture radius is expected to be a fraction of the size of the molecular motor, i.e., $\sim 10\,$nm. This is three to four orders of magnitude smaller than the typical filament lengths that are used in MP assays. On the other hand the longest relaxation time of a semiflexible filament of length $L$ varies as $\sim L^4$~\cite{Rubinstein2003}. To keep the calculations tractable, we choose a capture radius $r_c = 0.5\,\s$, smaller than the unit of length in the model, to be qualitatively consistent with the fact that this should be the shortest length scale of the problem. 
The active forces associated with MPs are known to be larger than that coming from thermal fluctuations, and 
we use $f_s = 2\, \kb T/\s$, $f_d = f_s$.  
The coarse-grained nature of the polymer segments considered allows multiple MPs to get associated with them, captured by our somewhat large MP density  in the 2d assay, $\rho = 3.8\, \s^{-2}$. 
The large spring constant $A = 100\, \s^{-1}$ is chosen to keep the bond length fluctuations small (within $5\%$). In absence of direct measurement of effective spring constant of active MPs (we are not considering the rigor bonds), we have chosen $k_m = A/\s$ for simplicity.
The attachment\,(detachment) of MP tails are stochastic, and performed using probabilities $\on\,\d t$($\off\, \d t$). The extension in the attached state has two contributors -- the MP tail is dragged along with the filament segment to which it is attached, and it can slide from one segment to another with an active velocity $v_t^a$. We study the influence of the active bed of MPs on the static and dynamic properties of the polymer as we vary the (a) bare processivity {$\Omega = \on/(\on + \omega_0)$ and (b) a dimensionless P{\'e}clet number defined as $Pe = v_0\sigma/D$. The numerical integrations are performed using $\d t = 10^{-3} \t$ for $Pe=1$, and $\d t = 10^{-4} \t$ for $Pe=10,\, 100$. Unless stated otherwise, we use $\o=5/6$, which corresponds to Kinesin MP property $\on:\w_0 = 5:1$~\cite{Leduc2004, Block1990, Vale1996}. The simulations are done over $2 \times 10^9$ steps,  and the steady state measurements are presented over $10^6$ configurations separated by $10^3$ steps, discarding the first $10^9$ steps. 
 
\section{Results}
\label{results}
At equilibrium,  mechanical and structural properties of a semiflexible filament are determined by the persistence ratio $u = L/\l$, where $L$ is the contour length of the chain, and $\l= 2 \k/(d-1)$ is the persistence length, where $d=2$ is the dimensionality of the embedding space~\cite{Dhar2002}.    
 The active drive from processive MPs attaching (detaching) to (from) the filament generates non-equilibrium stress which have profound effect on the steady state conformational properties of the polymer. To characterize the conformational properties, we obtain probability distribution of the end-to-end distance, $P(r,L)$, of the polymer. At equilibrium, this has the scaling form, $P(r,L) = \frac{1}{L^d}p(r/L,L/\lambda) = \frac{1}{L^d}p(\tilde r,u)$ where $\tilde r = r/L$ and $u = L/\lambda$. The limits of $u\to 0$ and $\infty$ are the limits of rigid rod and flexible polymers, respectively. 
 For equilibrium worm like chain,  $p(\tilde r,u)$  shows a first-order-like transition from a single maximum at $\tilde r = 0$ for the flexible limit of large $u$ to a maximum at $\tilde r = 1$ for a very rigid polymer with small $u$~\cite{Dhar2002, Chaudhuri2007}. 
 We choose the value of $u=3.33$, in the regime between these two limits where semiflexibility is most strongly pronounced~\cite{Chaudhuri2007}, to examine the impact of active MP bed on semiflexible polymers. In {\em in vitro} experiments, the ratio $u$ may be tuned by controlling persistence length of the chain by, e.g., changing salt concentration in the medium thereby changing interaction, or by stabilising the chain lengths. In all our simulations, unless stated otherwise, for $L=63\,\s$ chains, $u = 3.33$ sets $\l = 18.92\, \s$. Two typical configurations of the MP driven polymer is shown in Fig.\ref{figschem}($i$) and ($ii$) for $Pe=100$ and $\o=5/6$.

\subsection{How far from equilibrium the polymer morphology is\,?}
Under the active drive of the gliding assay of MPs, the morphology of the polymer changes. In Fig.\ref{figKL} we show how this impacts the end-to-end distribution function $p(\tilde r, u)$. The conformational change with respect to the equilibrium is well captured by the logarithmic ratio,
\bea
\D \S= \ln\left[ \frac{p(\tilde r)}{p_{eq}(\tilde r)}\right].
\eea
In Fig.\ref{figKL}($a$) we show how the dimensionless quantity 
$\D \S$ changes with activity. For comparison, the equilibrium distribution $p_{eq}$ is shown in  Fig.\ref{figKL}($c$).

If the activity of MPs is independent of the load force acting on them, $\off=\w_0$ and $v_t^a = v_0$. This corresponds to the limit of infinitely large $f_d$ and $f_s$. It is expected that the deviation $\D \S$ would be large for large non-equilibrium driving, quantified in terms of $f_d$, $f_s$ and $\o$. In Fig.\ref{figKL}, we explore the impact of activity using the moderate value of $Pe=1$.

We first consider the situation in which $\off=\w_0$ is kept fixed so that $\Omega = 5/6$, and the active velocity $v_t^a$ is varied\,(Fig.\ref{figKL}($a$)) for three possible situations. 
($i$)\,In the absence of any directed motion of the polymer, i.e., with $v_0 = 0$, $\D \S$ shows a dip near $\tilde r = 0$, indicating a relative bias to the {\em open} conformations of the polymer. This indicates that a mere stochastic attachment/detachment kinetics of MPs, that does not obey detailed balance, leads to an enhancement of effective stiffness of the filament. ($ii$)\,When attached, MPs move, and if the active velocity is assumed to be {\em independent} of the load experienced, we use $v_t^a = v_0$. The effect is dramatic. The filament, gliding on the attached MPs, undergoes a transition to a rotating spiral configuration (discussed further in Sec.~\ref{sec:coex}). This gives rise to a peak in $\D \S$ near $\tilde r = 0.1$. ($iii$)\,If we incorporate local stress dependence in $v_t^a$, the polymer is still softened but now switches between gliding and spiral states more freely. Thus in addition to the peak near $\tilde r = 0.1$, a non-zero value at higher $\tilde r$ appears in $\D \S$.  The statistics, dynamics and mechanical properties of the polymer under MP drive is determined by a competition between processive active velocity of MPs and bending stiffness of the polymer.

We next consider the situation allowing the detachment rate $\off$ to be {\em dependent} on the load force felt by individual MPs\,(Eq.\ref{eq:off}).
Given their similarity with the equilibrium distribution, $\Delta\Sigma \approx 0$ (Fig.\ref{figKL}($c$)), the corresponding non-equilibrium end-to-end distributions are shown explicitly in Fig.\ref{figKL}($b$). 
 The non-equilibrium stress build-up due to activity is relaxed easily by enhanced unbinding rate of stretched MPs allowing the polymer morphology to adopt equilibrium-like conformations. The distribution is closest to equilibrium for $v_t^a=0$. The strongest non-equilibrium feature is observed at stress {\em independent} activity $v_t^a=v_0$. At this point the distribution clearly shows a bi-modality with two maxima at $\tilde r \approx 0, 0.8$. Consideration of the stretching dependent decrease of active velocity as in Eq.\ref{velc}, decreases the height of the flexible chain maximum at $\tilde r \approx 0$, as the polymer  switches between gliding and spiral states more easily.

\begin{figure}[t]
\includegraphics[width=8cm]{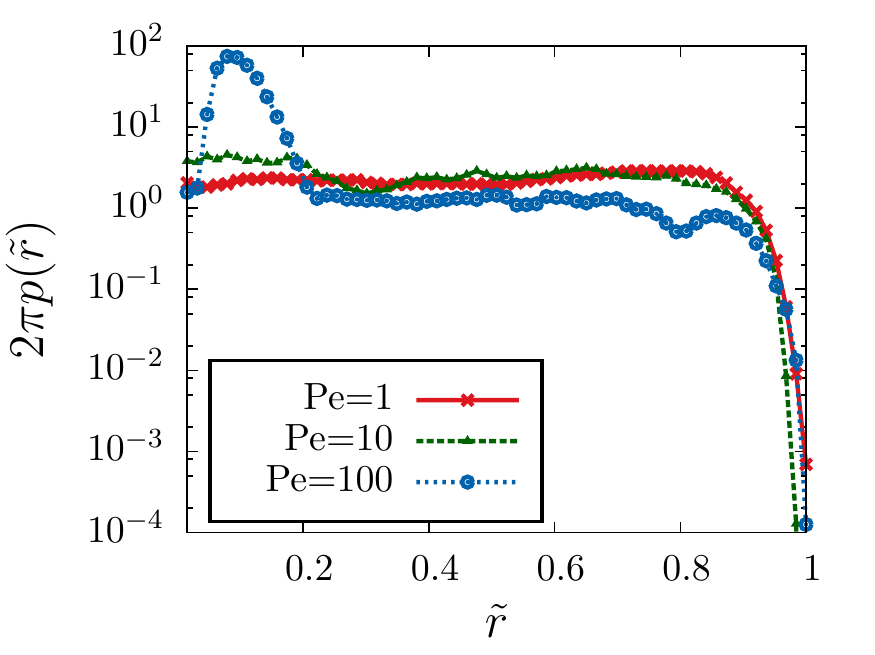}
\caption{(color online) End-to-end distribution for three different values of $Pe$ using stretching dependent turnover $\off$.  All other parameters are as in Fig.\ref{figKL}. 
}
\label{figPe}
\end{figure}

\begin{figure*}[t]
\begin{center}
\includegraphics[width=18cm]{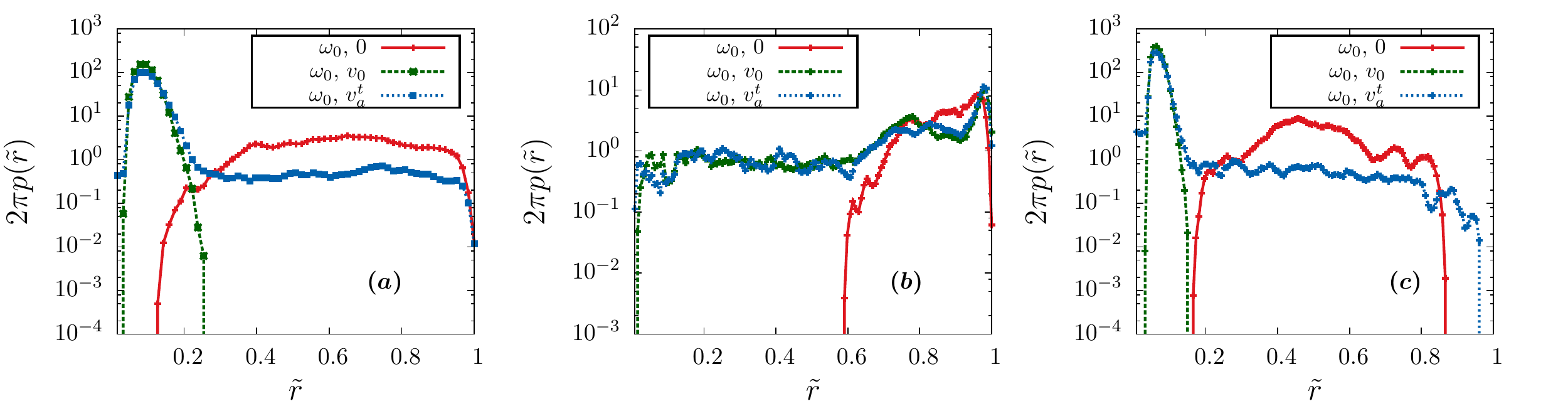}
\caption{(color online)~End-to-end distribution functions. We use constant detachment rate $\w_0$ with $\o=5/6$ for all the figures. The variation of MP active velocities are as in Fig.\ref{figKL}, with the non-zero active velocities set by $Pe=1$. The three graphs show results for ($a$) $N=64$, $u = 3.33$, $\l = 18.92\, \s$. ($b$) $N = 128$,  $u = 3.33$, $\l = 38.14\,\s$ ($c$) $N = 128$,  $u = 6.66$, $\l = 18.92\, \s$.} 
\label{figKappa}
\end{center}
\end{figure*}

In most biologically relevant situations, both the turnover and active motion of individual MPs depend on their instantaneous extension.  The activity is most strongly reflected in terms of the bare velocity of MPs $v_0$. As was shown in Ref.\cite{Schnitzer2000}, this velocity of unloaded Kinesin MP increases from $1\,$nm/s to finally saturate to $\sim 1\,\mu$m/s, as the ambient ATP concentration increases from $1\,\mu$M to $1\,$mM. The change in $v_0$ is captured by changing $Pe$ in our current setup.
In Fig.\ref{figPe}, we show how polymer properties vary with increasing $Pe$ when both $\off$ and $v_a^t$ are treated as local strain dependent quantities.
For low values of $Pe$, the local forces acting on the polymer backbone due to binding kinetics and motor movement is not sufficient to cause significant local curvature. As $Pe$ is increased, due to tangential velocity of MPs and enhanced directional fluctuations, the polymer starts to coil up and rotates with a spiral configuration in the steady state (discussed further in Sec.~\ref{sec:coex}). The impact shows up in terms of a maximum in $p(\tilde r)$ near $\tilde r=0.2$ appearing for large Peclet, $Pe=100$ (Fig.\ref{figPe}).
This feature is robust with respect to change in $\Omega$ (see Appendix-\ref{app:A}).

\subsection{Competition between activity and bending stiffness}
\label{app:B} 
For a semiflexible polymer in equilibrium, the end-to-end distribution $p(\tilde r,u)$ is determined by the dimensionless ratio $u=L/\l$. On the other hand, in presence of motor proteins,  the statistical and mechanical properties are expected to be determined by an interplay of activity and bending rigidity.  
To probe that within our model, here, we fix $\off = \omega_0$, and vary the chain length $L = (N-1)\sigma$ by changing $N$, the ratio $u=L/\l$, and persistence length $\l$ of the polymer to study their impact on conformational properties. We use both stress dependent and independent $v_a^t$, and plot the end-to-end distributions for three different active velocities in Fig.~\ref{figKappa}. A comparison of Figs.~\ref{figKappa}($a$) and ($b$) clearly shows that for the same $u$ and different $L$, unlike in equilibrium semiflexible chains, the conformational properties of the polymer are significantly different. For example, for $N = 128$ and $u = 3.33$ (Fig.~\ref{figKappa}($b$)), the distribution for $v_0 = 0$ indicates a much stiffer polymer compared to $N = 64$ (Fig.~\ref{figKappa}($a$)). 
For non-zero active velocity, the spiral states observed for $N = 64$ disappears for $N = 128$, leading to stiffer conformations devoid of spirals. If, however, we keep the value of $\l$ fixed as we change the length of the polymer from $N = 64$ to $N = 128$ (Fig.~\ref{figKappa}($c$)), the distributions we get compares much better with Fig.~\ref{figKappa}($a$). This suggests that, for a given processivity $\o$, the conformational properties of polymers driven by MPs are determined by a competition between active velocity and bending rigidity, and not by the ratio $u$. 

Within active polymer models with constant tangential drive, arguing that active force $f_p$ may generate compression, a torque balance leads to a critical active force $f^c_p \sim \l/L^3$, beyond which straight filaments are unstable towards buckling~\cite{Chelakkot2014a}. 
In the limit of stress independent activity, a simple extension of this relation to the instability of the filament under MP driving can be obtained by replacing $f^c_p = \a \o v^c_0$. This leads to a relation $v^c_0 \sim \l/\a \o L^3$. Thus buckling instabilities are expected to be controlled by the dimensionless number ${\cal F} = \a \o v_c L^3/\l $. However, for polymers driven by real MPs that shows stress dependent activity and turnover, the determining factors turn out to be more subtle.

\subsection{Determination of effective stiffness}
To further characterize the steady state conformational properties of the polymer, we consider the tangent-tangent correlation function, $\la \tv (s) \cdot \tv (s^{\prime})\ra$ for different $Pe$. 
For an equilibrium worm like chain, one expects a single  exponential decay of the correlations, characterized by the persistence length $\lambda$ as, $\la \tv (s)\cdot \tv (s^{\prime})\ra = \exp({-|s - s^{\prime}|/\l_{\rm eff}})$. In the long separation limit, the presence of self-avoidance leads to an effective power law correlation function determined by the Flory exponent, a behavior we ignore for relatively short length scales in the ensuing discussion. This results in a $\l_{\rm eff}$ that is larger than $\l$ in equilibrium simulations. The tangent-tangent correlation provides a measure for structural rigidity of the filament and can be determined from experiments by fluorescent imaging of polymer conformations.  
In Fig.~\ref{figMode}, we observe that the correlation function for small activity, $Pe=1$, shows a characteristic exponential decay that follows the equilibrium correlation function very closely. Fig.~\ref{figMode} shows that the correlation length decreases with increase in $Pe$. This is indicative of a softening of the polymer with the emergence of strong bending fluctuations. Up to $Pe=10$ shown in the graph, the overall nature can be described by a single exponential decay, which is fitted to extract the effective persistence length $\lambda_{\textrm{eff}}$, directly. For higher values of $Pe$, e.g., at $Pe=100$, the correlations start showing oscillations, capturing emergence of spiral conformations that occur at higher activity. In such cases, the crossing of zero by the correlation function is interpreted as the persistence length. The variation of this effective persistence length with activity is listed in Table-\ref{stiff}.
\begin{figure}[t]
\includegraphics[width=8cm]{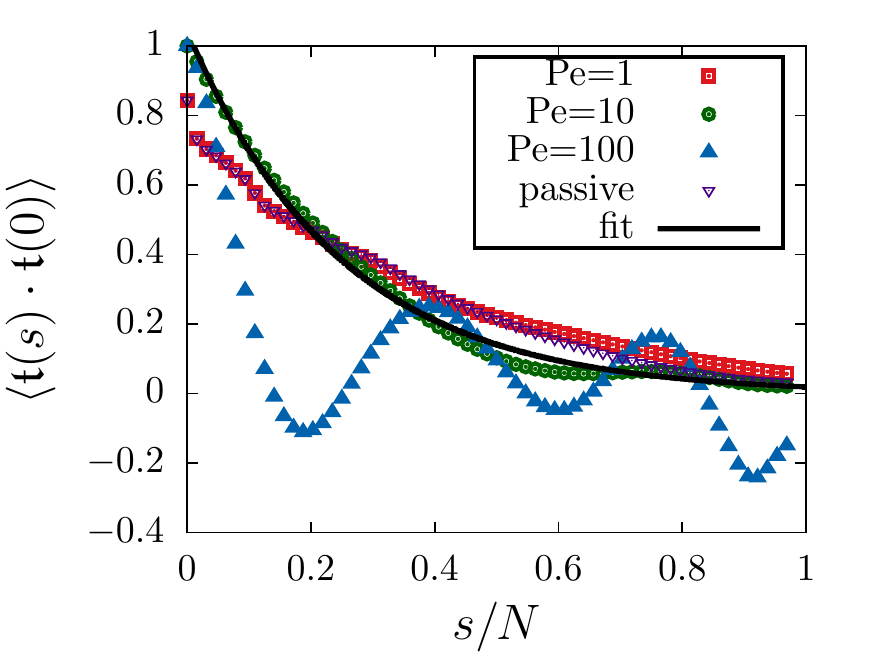}
\caption{(color online) Tanget-tangent correlation function for a chain of $N = 64, u = 3.33$, and activity $v_t^a$ set by $Pe = 1, 10$ and $100$ and load dependent detachment rate $\off$ with $\Omega = 5/6$. The data set {\em passive} denotes equilibrium result. The solid line shows a single exponential fit to $Pe=10$ data used to extract the effective persistence length $\l_{\rm eff}= (15.99\pm 0.24)\,\s$. 
}
\label{figMode}
\end{figure}
\begin{table}[h]
\small
	\caption{\ Activity modulated effective persistence length of a chain of length $L=63\,\s$ (with $N = 64$) and $\l = 18.92\, \s$. The table shows $\l_{\rm eff}$ obtained from tangent- tangent correlation function. With $Pe$, the persistence length first increases, and then decreases. 
   }
  \label{stiff}
\begin{tabular*}{0.48\textwidth}{@{\extracolsep{\fill}}ll} 
\hline
$Pe$ &  $\l_{\rm eff}/\s$ \\
\hline
{\rm equilibrium}  & $23.59 \pm 0.39$\\
\hline
$1$  & $25.21 \pm 0.23$\\
\hline
$10$  & $15.99 \pm 0.24$ \\
\hline
$100$  & $8.89$\\
\hline
\end{tabular*}
\end{table}

\begin{figure}[h]
\includegraphics[width=9cm]{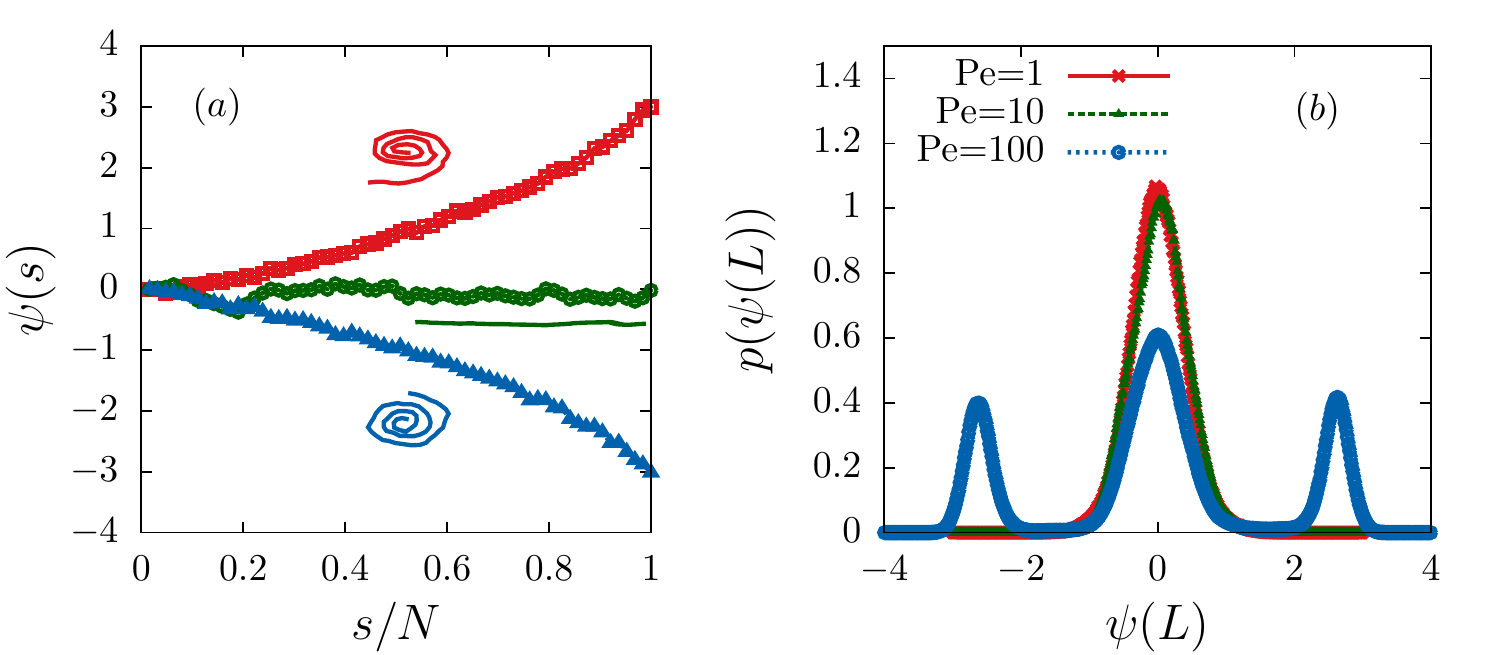}
  \caption{(color online) 
  Analysis of the turning number $\psi(s)$ using $N = 64, u = 3.33$ and load dependent detachment $\off$ with $\o = 5/6$ with activity $v_t^a$ set by $Pe$.
  ($a$)\,Plot of $\psi(s)$ for three different configurations with $Pe = 100$. It shows that $\psi(s)$ is an effective order parameter, distinguishing between the open state (green), clockwise spiral (blue) and anti-clockwise spiral (red). ($b$)\,Probability distributions for $\psi(L)$ at $Pe = 1, 10$ and $100$.
}
\label{figProb}
\end{figure}

\subsection{Coexistence of spiral and open chains} 
\label{sec:coex}
In order to quantify the observations of the different conformational states of the polymer, we use the turning number~\cite{Krantz1999}, $\psi(s) = (1/2\pi) \int_0^s ds'\,(\p \vartheta/\p s')$ where $\vartheta(s)$ is the angle subtended by the unit tangent $\hat t(s)$ with $x$-axis. 
This $\psi(s)$ is a good order parameter, clearly distinguishing  an open polymer from a spiral one and also separates clockwise and anticlockwise spiral states (Fig.~\ref{figProb}(a)). The steady state probability distribution of $\psi(s = L)$ is a Gaussian with a peak at $\psi(L) = 0$ for small $Pe$, indicating the absence of spiral states. Increasing $Pe$ has a dramatic effect on the distribution, with symmetric peaks emerging for non-zero $\psi(L)$ indicative of coexisting spiral states with equal probabilities of clockwise and anticlockwise winding, along with the open state characterized by $\psi(L)=0$. Such  phase coexistence is a characteristic feature of a non-equilibrium first order phase transition.
Similar coexistence of spiral and open conformations were observed earlier in an active polymer model characterized by constant tangential force~\cite{Isele-Holder2015}. It was not {\em a priori} clear that our current model would give rise to a similar conformational behavior, given that the activity in our model gets modified by the build up and release of local strain via load dependent activity and turnover. As we have already shown, in fact, the effect of local strain dependence reflects strongly in the end-to-end distribution functions $p(\tilde r)$. Further, as we show in the following section, this implies dynamical crossovers in mean squared displacement that are unlike the active polymer model.

\subsection{Anomalous dynamics of  the center of mass}
\label{sec:crossovers}

In Fig.~\ref{figCom}(a) we show mean squared displacement (MSD) of the polymer center of mass as a function of time, for three different $Pe$ values that are separated over two decades. At very short time scales the MSD shows an approximate {\em ballistic} scaling $\la \D r^2_{cm} \ra \sim t^2$ up to $t \approx 1$ at all $Pe$. With increasing time, five crossovers at $Pe=1$ can be clearly seen, these include three ballistic-diffusive crossovers and two diffusive-ballistic crossovers. At $Pe=10$, numerical integration required a smaller step size restricting the results to a shorter total time $t$. Otherwise, all the crossovers are retained at $Pe=10$, with a reduction in crossover times. The qualitative behavior  changes  as the activity is increased to a larger value,  $Pe=100$. At this regime the first ballistic-diffusive crossover almost vanishes.  At $t\approx 1$ one finds a barely discernible change in the slope which quickly gets back to ballistic scaling. This is due to an effective merger of the first diffusive-ballistic crossover to the first ballistic-diffusive one. 
The ballistic-diffusive crossovers discussed in this section is a recurring feature of active systems~\cite{Peruani2007, Selmeczi2005, Isele-Holder2015, Ghosh2014}. It is known that a persistent random walker undergoes a crossover from initial ballistic to a final diffusive motion, while directed random walkers show a crossover from short time diffusive to long time ballistic scaling~\cite{Peruani2007}.  In the following section we present a detailed explanation of the  crossovers observed. 
 
In Fig.~\ref{figCom}($b$), we show time evolution of the center of mass position of the polymer at $Pe=100$, indicating its various conformations associated with the trajectory. As the polymer takes a folded conformation, which is often a spiral in our system, the force generated in different segments by the gliding assay cancel each other, and the net directed force on the center of mass is negligible. As a result, the center of mass moves diffusively, getting mostly localized in a narrow region, albeit with an enhanced diffusivity. 
When the polymer retains a more open conformation, the gliding assay indeed generates directed force on the center of mass, leading to a ballistic motion over such periods shown by long directed trails. 

More quantitatively, the ballistic- diffusive crossovers are associated with changes in the evolution of the end-to-end extension $r_{ee}$, the orientation of the end-to-end vector $\phi$, and the root mean squared (RMS) fluctuation of the center of mass position $\sqrt{\D r_{cm}^2}$ along a single trajectory. In Fig.~\ref{figCom}(c) we show this at $Pe=100$. Clearly there are time-spans over which $r_{ee}$ remains close to zero, i.e., the polymer remains in a folded (spiral at $Pe=100$) state, e.g., between $t\approx 4.5 - 5 \times 10^5\, \t$. 
It should be noted that the formation of spiral happens at high $Pe$ as was shown in Sec.~\ref{sec:coex}. However, even at smaller $Pe$, the chain switches between open and non-spiral folded conformations. Non-spiral folds show a little higher value of $r_{ee}$ than when spirals form. There are other time windows over which $r_{ee}$ fluctuates rapidly between open and spiral states (e.g., between $t\approx 0 - 4 \times 10^5\, \t$). 

 As is shown in Fig.~\ref{figCom}(c), $\phi$ changes ballistically on a timespan over which $r_{ee}$ remains close to zero in a spiral state.  In particular, between $t= t_1$ and $t_2$ the spiral rotates {\em clockwise} ballistically reflected in a linear change in $\phi$ with a negative slope. During such time spans, the $r_{ee}$ of spirally folded polymer remains small, and the center of mass position of the polymer does not change appreciably, as is shown by the flat segment of $\sqrt{\D r_{cm}^2}$ in Fig.~\ref{figCom}(c) in this time-window. In the window of $t=t_2$ and $t_3$ the filament opens up switching between relatively close and open conformations stochastically captured by the strong fluctuations in $r_{ee}$. In such a state the directed rotation practically stops, captured by the flat, approximately parallel to $t$-axis portion of the $\phi(t)$ curve. The polymer encounters directed drive from MPs during the time-spans over which it opens up leading to  appreciable displacement $\sqrt{\D r_{cm}^2}$ of the centre of mass. Between $t= t_3$ and $t_4$, the polymer folds back into a spiral state again, and starts rotating in the {\em anti-clockwise} direction this time, captured by the linear increase in $\phi$, associated with characteristic flat segments of $r_{ee}$ and $\sqrt{\D r_{cm}^2}$.

\begin{figure*}[t]
\includegraphics[width=18cm]{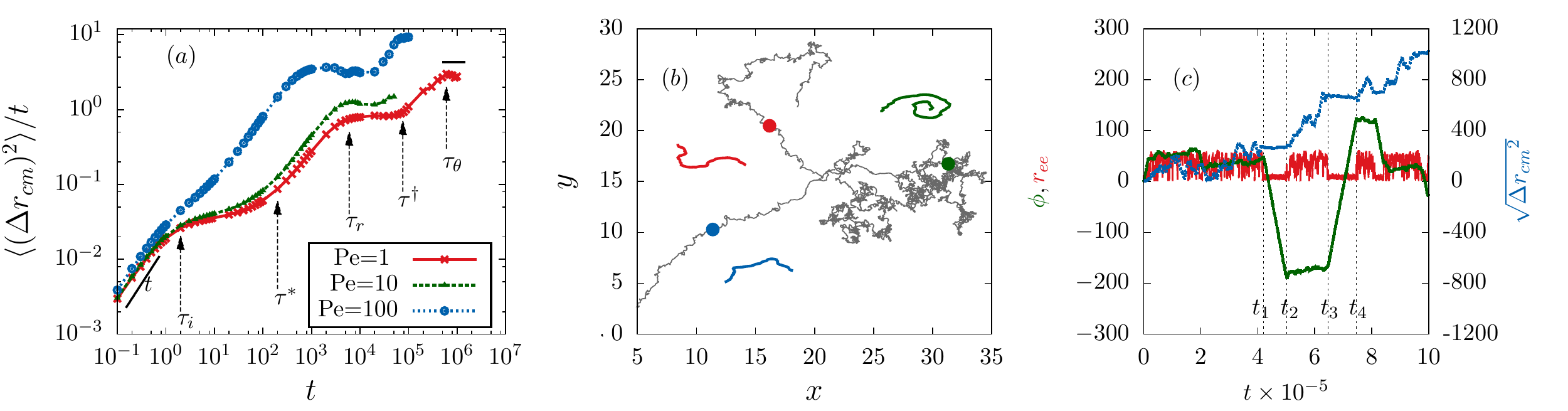}
	\caption{(color online) Dynamics of center of mass for a chain of $N = 64, u = 3.33$, with load dependent MP activity $v_t^a$ controlled by $Pe$ and detachment rate $\off$ determined by $\Omega = 20/21$. ($a$)\,Mean squared displacement of the center of mass at different P\'eclet ($Pe = 1, 10, 100$). Numerical analysis of the dynamics at $Pe=100$ is presented in ($b$) and ($c$). ($b$)\,The gray line shows a center of mass trajectory. Structure of polymer corresponding to the blue, red, and green points indicated on the trajectory are shown in the respective colors. ($c$)\,The end-to-end length $r_{ee}$ (red line), end-to-end orientation $\phi$ (green line), and root mean squared fluctuations of the center of mass position (blue line) for a single trajectory are shown as a function of time at $Pe=100$.}
\label{figCom}
\end{figure*}

\subsection{Ballistic to diffusive cross overs}
\label{crossover} 
To analyze the crossovers of the centre of mass MSD, let us first consider the dynamics of a particle in a Langevin heat bath in absence of any active drive, $m\, dv/dt = - \a v + \eta(t)$ where the Gaussian random noise obeys $\la \eta(t) \ra=0$ and $\la \eta(t) \eta(0) \ra = 2 \a \kb T \d(t)$. The corresponding displacement fluctuation of passive origin is given by 
\bea
\la \D r_{p}^2(t) \ra = 6 \f{\kb T}{m} \t_I^2 \left[ \f{t}{\t_I} - 1 + e^{-t/\t_I}\right],
\label{msd_eq}
\eea
where $\t_I = m/\a$.
For time scales $t \ll \t_I $ this leads to a ballistic scaling of MSD, $\la \D r_p^2(t) \ra  \approx 3 v_{eq} t^2$, with a velocity $v_{eq}=(2 \kb T/m)^{1/2}$.  At longer times $t \gtrsim \t_I$,  this crosses over to a diffusive scaling $\la \D r_p^2(t) \ra = 6 D_{eq} t$ with $D_{eq} = \kb T/\a$.  As is shown in Fig.\ref{figCom}($a$), the polymer centre of mass shows such a crossover near $\t_I = 1$ in our simulations~\footnote{Given that both mass and viscous friction scales similarly with the chain length, $\t_I = 1$.}.   

Because of the molecular motor drive, further ballistic-diffusive crossovers beyond $t_I$ are observed.  Following Ref.~\cite{Peruani2007}, we identify two possible mechanisms related to activity, ($i$)~the persistence of the direction of centre of mass velocity described by the correlation time $\t_\h$, and ($ii$)~the correlated fluctuations of the speed of the centre of mass with correlation time $\t_s$.  The speed fluctuations are approximately captured by an exponential correlation $ C_{v_s} (t) = \la \d v_s(t) \d v_s(0) \ra / \la \d v_s^2\ra \approx \exp(-t/\t_s)$ where $\d v_s = v_s - \la v_s \ra$~(Fig.~\ref{figvelcorr} in Appendix-\ref{app:C}). Similarly, the orientational fluctuation of velocity obeys $C_\h (t) = \la e^{i [\h(t) - \h(0)] } \ra \approx \exp(-t/\t_\h)$ where $\t_\h$ is the persistence time~(Fig.\ref{figthetacorr} in Appendix-\ref{app:C}). On the other hand, as we find, the velocity amplitude and orientations are only weakly correlated~(Fig.\ref{crosscorr} Appendix-\ref{app:C}). Such correlations can be ignored to use the expression of active displacement fluctuations,
\bea
\la \D r^2 (t) \ra &=& \la \D r_p^2(t) \ra + 2 {\la v_s \ra^2} \t_\h^2  \left( \f{t}{\t_\h} -1 + e^{-t/\t_\h}\right) \nn\\
 &&  + 2 \la \d v_s^2 \ra \t_r^2 \left[ \f{t}{\t_r} -1 + e^{-t/\t_r} \right], 
 \label{msd_active}
\eea
where, $\t_r^{-1} = \t_\h^{-1} + \t_s^{-1}$. 
In the above expression the speed $\la v_s \ra$ and its fluctuations $\la v_s^2\ra$ are due to activity controlled by $Pe$.   
If $\la \d v_s^2 \ra=0$, the above expression would suggest a ballistic dynamics for $t \ll \t_\h$, crossing over to diffusion at $t \gtrsim \t_\h$ as the direction of persistent motion diffuses. This is expected for structureless active Brownian particles with constant active speed.

However, in presence of speed fluctuations in the polymer, the other time-scale $\t_r < \t_\h$ intervenes. The total mean squared displacement of the polymer center of mass has contributions from both thermal fluctuations Eq.(\ref{msd_eq}) and activity Eq.(\ref{msd_active}).  If the three time-scales $\t_I \ll \t_r \ll \t_\h$ present in the problem are well separated, they are expected to lead to three ballistic-diffusive crossovers: ($a$)~At $t \ll \t_I$ one expects a ballistic motion $\la \D r^2_{cm} \ra \approx 3 v_{eq} t^2$ with a velocity $v_{eq} = (2 \kb T/m)^{1/2}$. ($b$)~At $t \gtrsim \t_I$ one crossover to diffusive regime takes place, with equilibrium diffusion constant $D_{\rm eq}=\kb T/\a$. This is the first ballistic-diffusive crossover, and is independent of activity. ($c$)~This regime lasts until $\t^\ast= 3 (v^2_{eq}/\la \d v_s^2\ra) \t_I$ at which the chain starts to respond to the active force that drives it in a directed manner. This gives rise to the first diffusive-ballistic crossover. For $\t^\ast < t \ll \t_r$, we find a ballistic behavior dictated by the active speed fluctuation $\sim \la \d v_s^2\ra t^2$.  A sufficiently strong activity can enhance $\la \d v_s^2\ra$ to reduce $\t^\ast$ to merge this active ballistic regime to the equilibrium ballistic scaling, as is seen for $Pe=100$ in our simulations. ($d$)~As $t$ crosses $\t_r$, the scaling of $\la \D r^2 (t) \ra$ crosses over to another diffusive regime, the second ballistic-diffsuive crossover, with effective diffusion constant $D \approx D_{\rm eq} + \f{1}{3} \la \d v_s^2 \ra \t_r$.   ($e$)~This regime persists until $\t^\dag = 3 (v_{eq}^2/\la v_s\ra^2) \t_I + 2 (\la \d v_s^2 \ra/\la v_s^2 \ra) \t_r$. Beyond this point the second diffusive-ballistic crossover takes place. For $\t^\dag < t \ll \t_\h$, the ballistic behavior is dictated by $\sim \la v_s \ra^2 t^2$. (f)~For $t \gtrsim \t_\h$, this ballistic regime slowly crosses over to the final diffusive behavior, the third ballistic-diffusive crossover, dictated by an effective diffusion constant   $D \approx D_{\rm eq} + \f{1}{3} (\la \d v_s^2 \ra \t_r + \la v_s \ra^2 \t_\h)$. This qualitatively explains the ballistic-diffusive crossovers obtained in Fig.\ref{figCom}($a$). 

Before ending this section, we note that, the fluctuations in active speed and orientation in the polymer arise essentially from the same driving mechanism due to molecular motors, and conformational relaxation of the polymer. Thus the two quantities may have similar fluctuations and significant cross-correlation. In Appendix-\ref{app:C} we show the auto-correlation functions of the center of mass speed and the active orientation, as well as the cross-correlation between the two. The auto-correlation data show longer correlation times for the orientational fluctuations. The cross-correlation function breaks time-reversal symmetry, capturing the non-equilibrium driven nature of the system, and shows  correlation even at long time gaps.

\section{Discussion and outlook}
\label{sec:outlook}
Using stochastic molecular dynamics simulations we have investigated the conformational and dynamical properties of a semiflexible polymer in the presence of motor proteins, which (un)bind (from)\,to the polymer and perform directed active motion. Unlike in the equilibrium worm like chain, the end-to-end statistics in this case is not controlled by the ratio of persistence length and chain length, but results from a local competition between the processive active velocity and bending rigidity. As is shown in this paper, local stress dependence of turnover and active velocity provides new relaxation mechanisms giving rise to steady states unlike the active polymer models with constant tangential self-propulsion.
The activity influences polymer morphology, mechanical properties, and dynamics in a concerted manner. With increasing activity of the motor proteins, we observed the following : ($i$)\,The end-to-end distribution characterising polymer conformation shows both stiffening and softening relative to the equilibrium morphology associated with the build up of local active stress and its relaxation. ($ii$)\,The stretching dependent active velocity and turnover of molecular motors gives rise to an interplay of three time scales, the inertial, orientational and speed relaxation times of the centre of mass, leading to a series of ballistic - diffusive crossovers in the mean squared displacement of the centre of mass. 

These crossover time scales can be interpreted into real times using dynamics of a filament of $\sim 2\,\mu$m length. For example, considering $\sigma = 20$ nm, the $64$ bead chain can be interpreted to have a length $1.84 \mu $m. This sets $f_s = f_d = 0.4$ pN, slightly smaller than the pN scale in, e.g., kinesin molecules.  Assuming the viscosity of ambient fluid $100$ times that of water, i.e., equivalent to that in cytoskeleton~\cite{Howard2005}, one obtains a viscous drag of $\alpha=0.02$ pN s$/\mu$m. This sets the unit of time $\tau=0.002\,$s. 
As expected, only the non-inertial time-scales are relevant from the perspective of slow dynamics. Interpreting the predictions from Fig.\ref{figCom}($a$) we find the following slow time scales, 
the ballistic-diffusive crossover times $\tau_r \sim 15$s, and $\tau_\theta \sim 15$ minutes, and the diffusive-ballistic crossover time $\tau^\dagger \sim 3$ minutes. 
The predictions presented here are amenable to verification in experiments on molecular motor assays.

While our system reproduces some of the predictions of the standard active polymer model, some other properties that we observe are entirely due to the strain dependence of the activity and turnover of MPs. For example, the observed activity dependent reduction of effective bending stiffness, and the coexistence of spiral  and open conformations at an activity beyond a critical value are expected within the active polymer model. On the other hand, the detailed nature of end-to-end distribution functions, and the series of ballistic-diffusive crossovers observed in the center of mass dynamics are features that are unlike active polymer models~\cite{Isele-Holder2015}.

\section*{Acknowledgements}
A.C and D.C. acknowledge SERB, India for financial support through grant number EMR/2014/000791. DC thanks SERB, India, for financial support through grant number EMR/2016/001454, and ICTS-TIFR, Bangalore for an associateship. N.G. acknowledges UGC, India for a fellowship.
%
%
\begin{figure*}[t]
\begin{center}
\includegraphics[width=18cm]{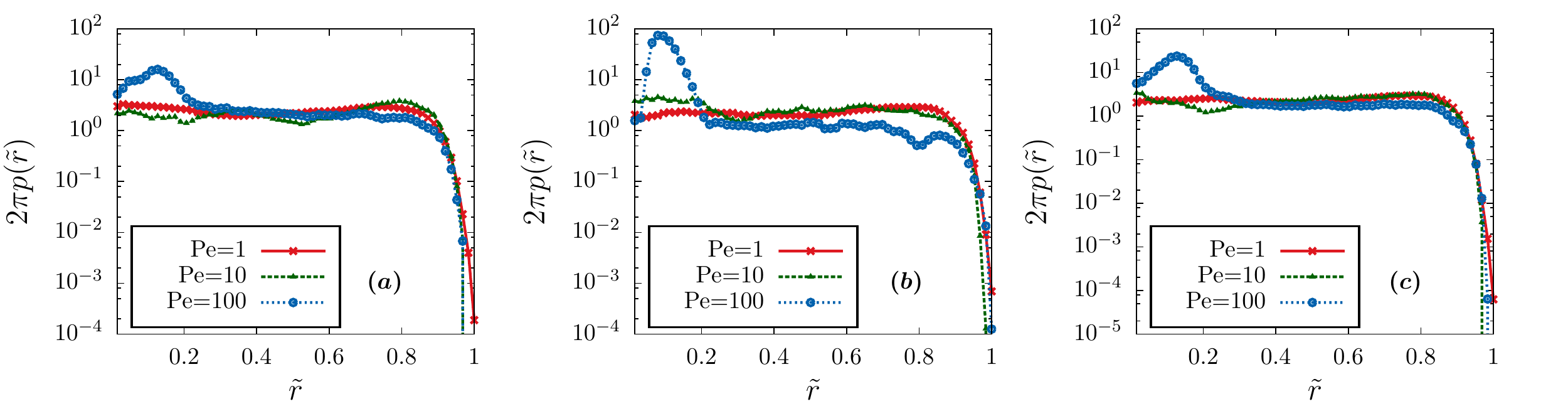}
\caption{(color online)~End-to-end distribution at $Pe=1,\,10,\,100$ for different values of bare processivity $\Omega$, using stress dependent active velocity and detachment rate with $N = 64, u = 3.33$. The graphs correspond to (a) $\Omega = 2/3$, (b) $\Omega = 5/6$, and (c) $\Omega = 20/21$.}
\label{figOmega}
\end{center}
\end{figure*}

\begin{figure}[t]
\begin{center}
\includegraphics[width=8cm]{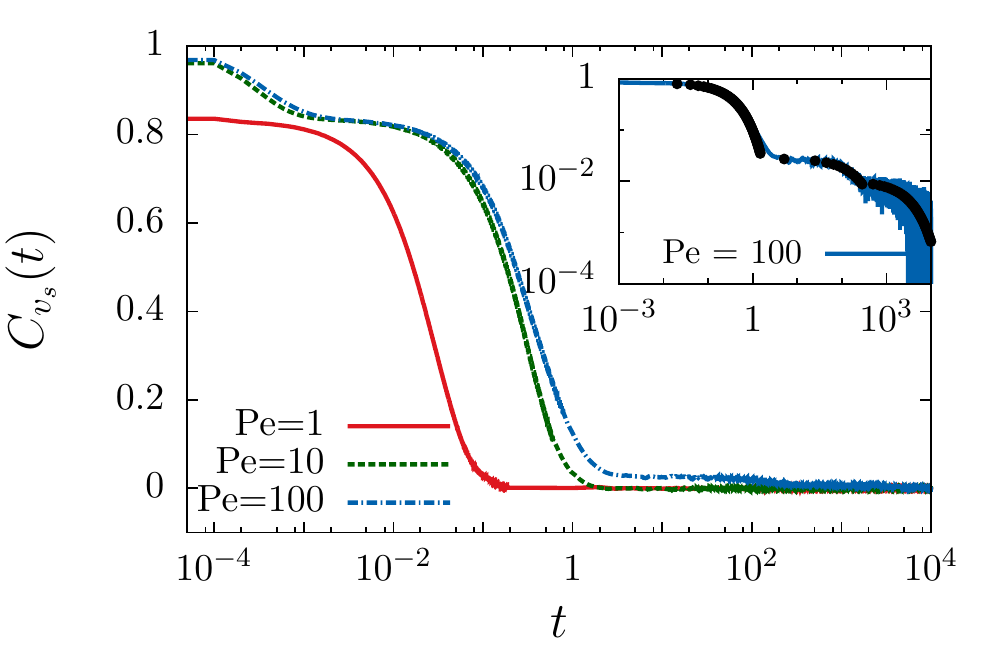}
	\caption{(color online)~Speed autocorrelation of the centre of mass of the polymer. Single exponential decays are observed for both $Pe = 1,\,10$, with correlation times $t_s \approx 0.1\tau,\,1.0\tau$ respectively. (Inset) In the log-log plot, for $Pe = 100$, multiple exponential decays with $t_s \approx 1\tau,\, 250\tau,\,2000\tau$ are shown. The three exponential fits are indicated by black points.}
\label{figvelcorr}
\end{center}
\end{figure}

\begin{figure}[t]
\begin{center}
\includegraphics[width=8cm]{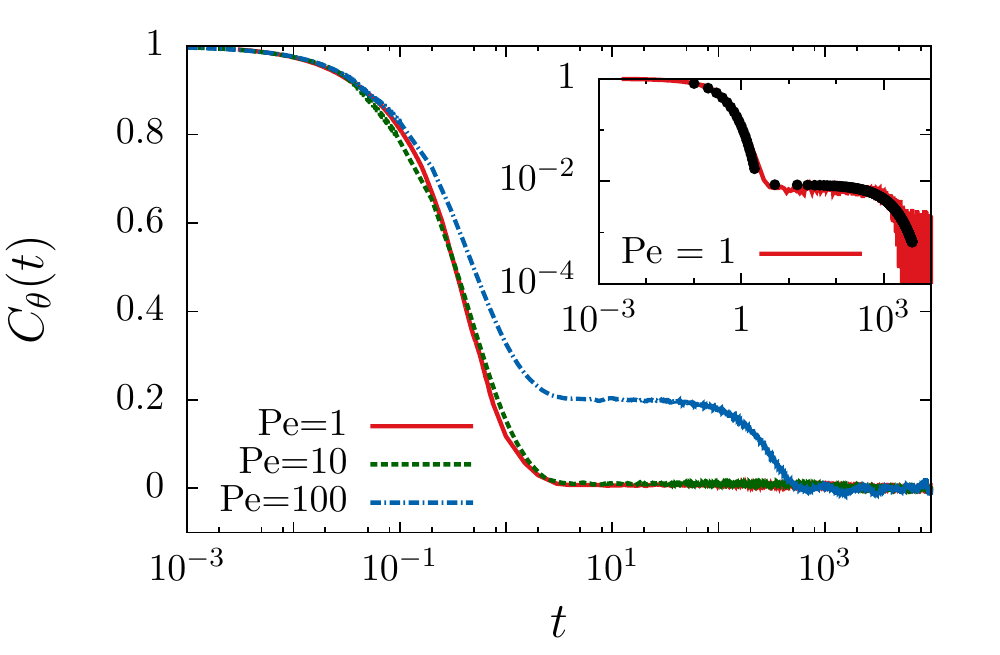}
	\caption{(color online)~Orientational autocorrelation of the centre of mass velocity vector. This shows multiple exponential decays for all $Pe$. (Inset) For $Pe = 1$, the log-log plot shows multiple exponential decays with time scales $t_{\theta} \approx 1\tau,\,1500\tau$. The two exponential fits are indicated by black points.}
\label{figthetacorr}
\end{center}
\end{figure}
\begin{figure}[t]
\begin{center}
\includegraphics[width=8cm]{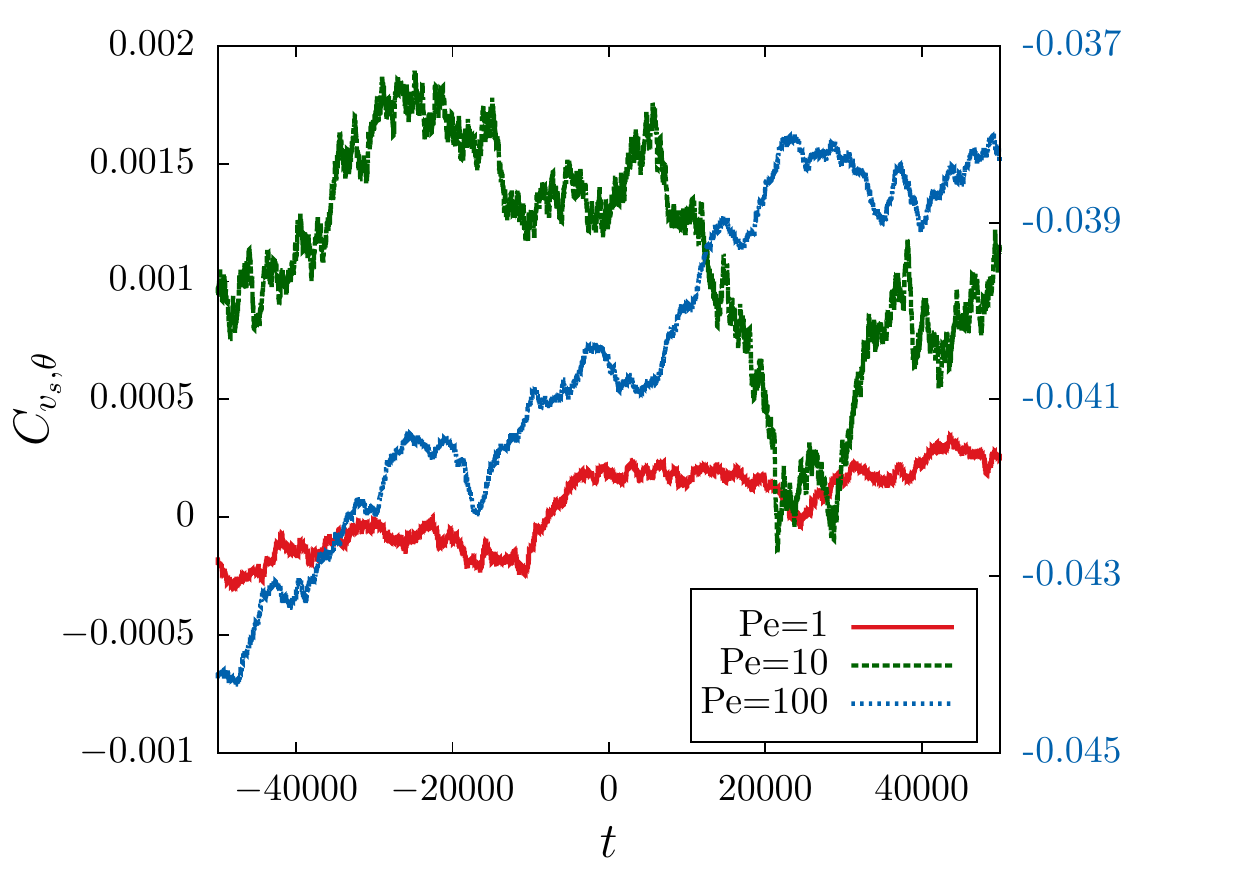}
	\caption{(color online) Cross-correlation of the orientation and speed of the centre of mass velocity for $Pe = 1$ (red), $10$ (green) and $100$ (blue, and values correspond to right ordinate).}
\label{crosscorr}
\end{center}
\end{figure}

\appendix

\section{End-to-end distributions for different $\Omega$}
\label{app:A}
In Fig.~\ref{figOmega}, we show the dependence of the conformational properties of the polymer as the bare processivity $\Omega=\on/(\on+\w_0)$ is varied. Here we consider the scenario where both the detachment rate and the active velocity depend on the local stress. For a fixed $\Omega$ we plot the end-to-end distribution of the polymer as $Pe$ is changed. As in Fig.~\ref{figKL}(c), the distributions look similar to equilibrium distribution $p(\tilde{r})$ for low $Pe$ and a peak near $\tilde{r} \approx 0$ appears for high $Pe$, indicating the emergence of spiral states. Therefore we conclude that for stress dependent $\off$, varying $\Omega$ does not affect the conformational properties significantly. Recall that a stress independent $\off$ with non-zero $Pe$ results in coiled states of the polymer. Switching on local stress dependence in $\off$ allows the polymer to relax back to its equilibrium conformations whenever stress builds up beyond a limit, even if the processivity $\Omega$ is  high. 
As $Pe$ is increased, it triggers an instability towards spiral states and we see the emergence of a peak near $\tilde r=0$ in the steady state distributions.

\section{Centre of mass dynamics}
\label{app:C}
In this section we analyse autocorrelation of the centre of mass velocity vector, focussing on the speed $v_s(t)$, and orientation $\h(t)$ separately. Here we distinguish between the direct measures of the correlation times $t_s$ and $t_\h$ associated with multi-exponential decays of correlations, from the assumptions of single exponential decays with $\t_s$, $\t_\h$ used in the analysis of dynamical cross-overs in Sec.~\ref{crossover}.  
In Fig.~\ref{figvelcorr} we show the auto-correlation of speed, $C_{v_s} (t) = \la \d v_s(t) \d v_s(0) \ra / \la \d v_s^2\ra$. 
 A fast single exponential decay $\exp(-t/t_s)$ is observed at both $Pe=1,\,10$, with $t_s\approx 0.1\t, 1.0\t$, respectively.  However, at $Pe=100$, we observe multiple exponential decays with time scales $t_s \approx 1\t,\, 250\t,\, 3600\t$ (see the inset of Fig.~\ref{figvelcorr}). 
 
 The orientational correlation $C_\h(t) = \la e^{i[\h(t)- \h(0)]}\ra$, shows multiple exponential decays at all $Pe$ values~(Fig.~\ref{figthetacorr}). The initial decay is fast with $t_\h \approx 1\t$. For $Pe = 1, 10$ we can extract the longer time scales, as shown in the log-log plot in the inset for $Pe = 1$, to give  $t_\h=1500\t, 2000\t$ respectively. However, for $Pe = 100$, in the absence of better averaging, it is difficult to extract the longest time scale.

 Moreover, the speed and orientations remain correlated. The cross-correlation functions $C_{v_s,\h}(t) = \la  v_s(t) \h(0) \ra/ [ \sqrt{\la \d v_s^2\ra} \sqrt{\la \d \h^2\ra} ] $ calculated for $Pe=1,\,10,\ 100$ are shown in 
 Fig.~\ref{crosscorr}. All of them show significant correlation, which remarkably do not decay with increasing time-gap. The asymmetry of the data around $t=0$ captures the break-down of time-reversal symmetry due to the non-equilibrium molecular motor drive.

\end{document}